\documentclass[smallextended]{svjour3}  
\smartqed  

\usepackage{threeparttable}
\usepackage{graphicx}
\usepackage{qcircuit}
\usepackage{dcolumn}
\usepackage{bm}
\usepackage{booktabs}
\usepackage{CJK}
\usepackage{subfigure}
\usepackage{multicol}
\usepackage{color}
\usepackage{braket}
\usepackage{listings}
\usepackage{amsmath}

\begin{document}

\title{Entanglement constraint on wave-particle duality for tripartite systems
}
\subtitle{}


\author{Z.J. Li \and Y.Q. He$^\dag$ \and D. Ding \and T. Gao$^\S$ \and F.L. Yan*
}


\institute{
           Z.J. Li \at
              College of Physics, Hebei Normal University, Shijiazhuang 050024, China\\
           \and
          $^\dag$ Y.Q. He \at
              College of Science, North China Institute of Science and Technology, Beijing 101601, China\\
              \email{heyq@ncist.edu.cn}\\
               \and
           D. Ding \at
              College of Science, North China Institute of Science and Technology, Beijing 101601, China\\
               \and
          $^\S$ T. Gao \at
              School of Mathematical Sciences, Hebei Normal University, Shijiazhuang 050024, China\\
              \email{gaoting@hebtu.edu.cn}\\
              \and
           * F.L. Yan \at
              College of Physics, Hebei Normal University, Shijiazhuang 050024, China\\
              \email{flyan@hebtu.edu.cn}
}

\date{Received: date / Accepted: date}

\maketitle

\begin{abstract}
A global multi-partite entanglement may place a constraint on the wave-particle duality. We investigate this constraint relation of the global entanglement and the quantitative wave-particle duality in tripartite systems. We perform quantum state tomography to reconstruct the reduced density matrix by using the OriginQ quantum computing cloud platform. As a result, we show that, theoretically and experimentally, the quantitative wave-particle duality is indeed constrained by the global tripartite entanglement.

\end{abstract}

\section{Introduction}
\label{intro}
The principle of wave-particle duality plays an important role in quantum physics.
In 1924, De Broglie \cite{Ph.D.thesis(1925)} first proposed the hypothesis of matter wave, which is intrinsically related to the Bohr complementarity principle \cite{Nature.121.580(1928)} and the Heisenberg uncertainty relation \cite{Z.Phys.33.879(1925)}. In recent years, with the development of quantum physics, especially quantum information theory, how to quantify the wave-particle duality has received the most attention.

In 1979, Wootters and Zurek \cite{PhysRevD.19.473(1979)} first investigated the problem of quantifying wave-particle duality. Soon afterward Greenberger and Yasin \cite{Phys.Lett.A.128.391(1988)} proposed a duality relation $P^{2}+V^{2}=1$ for the pure state of a single particle, where $P$ is path predictability characterizing particleness, and $V$ is the amount of interference describing waviness. Also, Englert \cite{Phys.Rev.Lett.77.2154(1996)} derived $D^{2}+V^{2}\leq1$, where $D$ denotes path distinguishability to quantify the particleness and $V$ is fringe visibility to quantify the waviness.
Since then, other interesting aspects of quantifying wave-particle duality have been investigated, such as asymmetric beams \cite{Proc.R.Soc.A.468.1065(2012),Phys.Rev.A.98.022130(2018)}, multi-slit interference \cite{Phys.Rev.A.64.042113(2001),Int.J.Quant.Inf.6.129(2008),Phys.Rev.A.92.012118(2015),Phys.Rev.A.93.062111(2016)}, etc.
These studies mainly focus on the case of the single particle.

For bipartite systems, it is found that one should consider three quantities $P$, $V$ and the entanglement measure between two particles \cite{Phys.Rev.A.48.1023(1993),Phys.Rev.A.51.54(1995)}. Typically, Jakob and Bergou \cite{Opt.Commun.283.827(2010)} have reported a triality relation for the bipartite system as $P^{2}+V^{2}+C^{2}=1$, where $C$ is the well-known bipartite entanglement measure, concurrence. So far, the research on the triality relation for bipartite systems has achieved fruitful results \cite{Phys.Rev.A.76.052107(2007),Phys.Rev.A.103.022409(2021),Opt.Lett.46.492(2021)}. For multi-partite systems, more recently, Marrou \cite{josaa.40.4.c22(2023)} et al. designed the experimental tests to show the triality relations using coupled photons involving the polarization and the spatial degrees of freedom. However, such relations were constrained only by the concurrence rather than the global tripartite entanglement.

In this paper, we focus on quantifying wave-particle duality constrained by the global tripartite entanglement.
Based on quantum state tomography, we design quantum circuit to test the constraint relation by using the OriginQ quantum computing cloud platform.

\section{Quantitative wave-particle duality}
\label{sec:1}

For a single qubit system, the quantum state can be represented by the density matrix
\begin{eqnarray}\label{}
\rho_{}=\left(
  \begin{array}{cc}
    \rho_{11} & ~\rho_{12}\\
    \rho_{21} & ~\rho_{22}\\
  \end{array}
\right).
\end{eqnarray}
Density matrix $\rho$ can be used to characterize waviness and particleness \cite{Phys.Rev.A.64.042113(2001)}, namely, visibility and predictability.

The visibility is usually defined by
\begin{eqnarray}\label{}
V=2|\rho_{12}|.
\end{eqnarray}
It is related to non-diagonal elements of the density matrix under two orthogonal bases $|0\rangle$ and $|1\rangle$, and it measures the waviness for the single qubit systems.
While for the particleness, it can always be measured by the diagonal elements of the density matrix, that is, the predictability
\begin{eqnarray}\label{}
P=|\rho_{22}-\rho_{11}|
\end{eqnarray}
introduced in the two-beam interferometers \cite{Phys.Lett.A.128.391(1988),Phys.Rev.Lett.77.2154(1996)}.

More generally, for multipartite systems, the waviness and particleness of the $k$-th subsystem obey the same rules as single-qubit \cite{Opt.Commun.283.827(2010),Phys.Rev.A.103.022409(2021)}. One can express them as
\begin{eqnarray}\label{}
V_{k}=2|\rho_{k_{12}}|
\end{eqnarray}
and
\begin{eqnarray}\label{}
P_{k}=|\rho_{k_{22}}-\rho_{k_{11}}|,
\end{eqnarray}
where $\rho_{k}$ is the reduced density matrix of the $k$-th subsystem.

\section{Tripartite entanglement constraint on wave-particle duality}
\label{sec:2}
Consider the tripartite systems consisting of three qubits, i.e., $A$, $B$ and $C$, and an arbitrary pure state of such systems can be expressed as
\begin{eqnarray}\label{E-LHV}
\begin{split}
|\psi_{ABC}\rangle =&a|000\rangle+b|001\rangle+c|010\rangle+d|100\rangle+e|110\rangle+f|101\rangle\ \\
&+g|011\rangle+h|111\rangle,
\end{split}
\end{eqnarray}
where $a,b,c,d,e,f,g,h$ are complex numbers satisfying $|a|^{2}+|b|^{2}+|c|^{2}+|d|^{2}+|e|^{2}+|f|^{2}+|g|^{2}+|h|^{2}=1$.
The density matrix of the state $|\psi_{ABC}\rangle$ is
\begin{eqnarray}\label{}
\rho_{ABC}&=|\psi_{ABC}\rangle\langle\psi_{ABC}|
\end{eqnarray}
and the reduced density matrices of the subsystems $A$, $B$ and $C$ are respectively
\begin{eqnarray}\label{}
\begin{split}
\rho_{A}&=\textrm{Tr}_{BC}(\rho_{ABC})\ \\
&=\left(
  \begin{array}{cc}
    aa^{*}+bb^{*}+cc^{*}+gg^{*}& ~ad^{*}+bf^{*}+ce^{*}+gh^{*}\\
    a^{*}d+b^{*}f+c^{*}e+g^{*}h& ~dd^{*}+ee^{*}+ff^{*}+hh^{*}\\
  \end{array}
\right),
\end{split}
\end{eqnarray}
\begin{eqnarray}\label{}
\begin{split}
\rho_{B}&=\textrm{Tr}_{AC}(\rho_{ABC})\ \\
&=\left(
  \begin{array}{cc}
    aa^{*}+bb^{*}+dd^{*}+ff^{*}& ~ac^{*}+bg^{*}+de^{*}+fh^{*}\\
    a^{*}c+b^{*}g+d^{*}e+f^{*}h& ~cc^{*}+ee^{*}+gg^{*}+hh^{*}\\
  \end{array}
\right)
\end{split}
\end{eqnarray}
and
\begin{eqnarray}\label{}
\begin{split}
\rho_{C}&=\textrm{Tr}_{AB}(\rho_{ABC})\ \\
&=\left(
  \begin{array}{cc}
    aa^{*}+cc^{*}+dd^{*}+ee^{*}& ~ab^{*}+cg^{*}+df^{*}+eh^{*}\\
    a^{*}b+c^{*}g+d^{*}f+e^{*}h& ~bb^{*}+ff^{*}+gg^{*}+hh^{*}\\
  \end{array}
\right).
\end{split}
\end{eqnarray}
We calculate $V_k$ and $P_k$, respectively, and we have
\begin{eqnarray}\label{V-A}
V_{A}&=2|ad^{*}+bf^{*}+ce^{*}+gh^{*}|,
\end{eqnarray}
\begin{eqnarray}\label{V-B}
V_{B}&=2|ac^{*}+bg^{*}+de^{*}+fh^{*}|,
\end{eqnarray}
\begin{eqnarray}\label{V-C}
V_{C}&=2|ab^{*}+cg^{*}+df^{*}+eh^{*}|,
\end{eqnarray}
\begin{eqnarray}\label{P-A}
P_{A}=||d|^{2}+|e|^{2}+|f|^{2}+|h|^{2}-|a|^{2}-|b|^{2}-|c|^{2}-|g|^{2}|,
\end{eqnarray}
\begin{eqnarray}\label{P-B}
P_{B}=||c|^{2}+|e|^{2}+|g|^{2}+|h|^{2}-|a|^{2}-|b|^{2}-|d|^{2}-|f|^{2}|
\end{eqnarray}
and
\begin{eqnarray}\label{P-C}
P_{C}=||b|^{2}+|f|^{2}+|g|^{2}+|h|^{2}-|a|^{2}-|c|^{2}-|d|^{2}-|e|^{2}|.
\end{eqnarray}

Then it is not difficult to check that
\begin{eqnarray}\label{}
\sum_{k=A,B,C}(V_{k}^{2}+P_{k}^{2}) \leq 3.
\end{eqnarray}
This means that these quantities, like $V_{k}$ and $P_{k}$, that are associated with the waviness and particleness of the subsystems can be constrained by such an inequality.
In order to build up a complete description of wave-particle duality for tripartite systems with equality rather than inequality, just as for bipartite systems \cite{Opt.Commun.283.827(2010)}, it is natural to think of the tripartite entanglement.

Here we consider the multi-partite entanglement measure \cite{J.Math.Phys.43.4273(2002),0305094} $Q=2[1-\frac{1}{n}\sum_{k=0}^{n-1}\textrm{Tr}(\rho_{k}^{2})]$, where $\rho_{k}$ is the reduced density matrix of the $k$-th subsystem of  the combined system $\rho$.
Let $n=3$, then for the state $|\psi_{ABC}\rangle$ we have
\begin{eqnarray}\label{}
Q=\frac{2}{3}[3-\textrm{Tr}(\rho_{A}^{2})-\textrm{Tr}(\rho_{B}^{2})-\textrm{Tr}(\rho_{C}^{2})].
\end{eqnarray}
A straightforward calculation using Eqs.(\ref{V-A})-(\ref{P-C}) shows that
\begin{eqnarray}\label{triality}
Q+\frac{1}{3}\sum_{k=A,B,C}V_{k}^{2}+\frac{1}{3}\sum_{k=A,B,C}P_{k}^{2}=1.
\end{eqnarray}
This leads to the conclusion that there exists a constraint relation of the global tripartite entanglement, quantitative waviness and particleness for the tripartite systems. Thus, one can argue that indeed the quantities associated with the waviness and particleness of the subsystems are constrained by the ``global entanglement''.

\section{Experimental test of the triality relation}

We next verify the present constraint relation for tripartite systems by means of quantum state tomography.
\subsection{Quantum state tomography}

Quantum state tomography \cite{Phys.RevA.64.052312.(2001)} is an effective method for determining the state of a quantum system.
It uses projective measurements on a collection of identically prepared systems and then it is capable of reconstructing the density matrix of the unknown quantum system.

For a single qubit, select a set of measurement operators $\mu_{0}=|0\rangle\langle0|+|1\rangle\langle1|$, $\mu_{1}=|0\rangle\langle0|$, $\mu_{2}=|-\rangle\langle-|$ and $\mu_{3}=|R\rangle\langle R|$, where $|-\rangle=(|0\rangle-|1\rangle)/\sqrt{2}$ and $|R\rangle=(|0\rangle-\textrm{i}|1\rangle)/\sqrt{2}$.
These four measurements are respectively related by the counts
$n_0=N/2(\langle0|\rho|0\rangle+\langle1|\rho|1\rangle)$, $n_1=N\langle0|\rho|0\rangle$, $n_2=N\langle-|\rho|-\rangle$ and $n_3=N\langle R|\rho|R \rangle$, where $N$ is a constant on the experimental configuration.
By this, it will suffice to reconstruct single qubit state as
\begin{eqnarray}\label{single_rho}
  \rho&=\frac{1}{2}\sum^{3}_{i=0}\frac{S_{i}}{S_{0}}\sigma_{i},
\end{eqnarray}
where $S_{0}=2n_{0}$, $S_{1}=2(n_{1}-n_{0})$, $S_{2}=2(n_{2}-n_{0})$ and $S_{3}=2(n_{3}-n_{0})$ are the Stokes parameters,
and $\sigma_{0}=|R\rangle\langle R|+|L\rangle\langle L|$, $\sigma_{1}=|R\rangle\langle L|+|L\rangle\langle R|$, $\sigma_{2}=\textrm{i}(|L\rangle\langle R|-|R\rangle\langle L|)$ and $\sigma_{3}=|R\rangle\langle R|-|L\rangle\langle L|$ are the Pauli matrices. Here $|L\rangle=(|0\rangle+\textrm{i}|1\rangle)/\sqrt{2}$.

\subsection{Experimental test using OriginQ quantum computer}

OriginQ offers a cloud platform to carry out quantum calculation with superconducting quantum computers \cite{OriginQ}.
We now turn to the experimental test of the constraint relation on the quantum cloud platform.

First, consider the states of tripartite systems
\begin{eqnarray}\label{3-thete-state}
  |\psi\rangle_{ABC}=&\cos\frac{\theta_{1}}{2}\cos\frac{\theta_{2}}{2}\cos\frac{\theta_{3}}{2}|000\rangle+\cos\frac{\theta_{1}}{2}\cos\frac{\theta_{2}}{2}\sin\frac{\theta_{3}}{2}|001\rangle \nonumber\\
  &+\cos\frac{\theta_{1}}{2}\sin\frac{\theta_{2}}{2}\cos\frac{\theta_{3}}{2}|010\rangle+\sin\frac{\theta_{1}}{2}\sin\frac{\theta_{2}}{2}\sin\frac{\theta_{3}}{2}|100\rangle \nonumber\\
  &+\cos\frac{\theta_{1}}{2}\sin\frac{\theta_{2}}{2}\sin\frac{\theta_{3}}{2}|011\rangle+\sin\frac{\theta_{1}}{2}\cos\frac{\theta_{2}}{2}\sin\frac{\theta_{3}}{2}|110\rangle \nonumber\\
  &+\sin\frac{\theta_{1}}{2}\sin\frac{\theta_{2}}{2}\cos\frac{\theta_{3}}{2}|101\rangle+\sin\frac{\theta_{1}}{2}\cos\frac{\theta_{2}}{2}\cos\frac{\theta_{3}}{2}|111\rangle,
\end{eqnarray}
which can be prepared by using the rotation operators
$R_{y}(\theta_{i})=\left(
  \begin{array}{cc}
    \cos{\theta_{i}}/{2} & -\sin{\theta_{i}}/{2}\\
    \sin{\theta_{i}}/{2} & \cos{\theta_{i}}/{2}\\
  \end{array}
\right)$, $i=1,2,3$, and two CNOT gates with the initial state $|0\rangle^{\otimes 3}$, as shown in Fig.1.

\begin{figure}[h]
      \centering
      \includegraphics[width=3in]{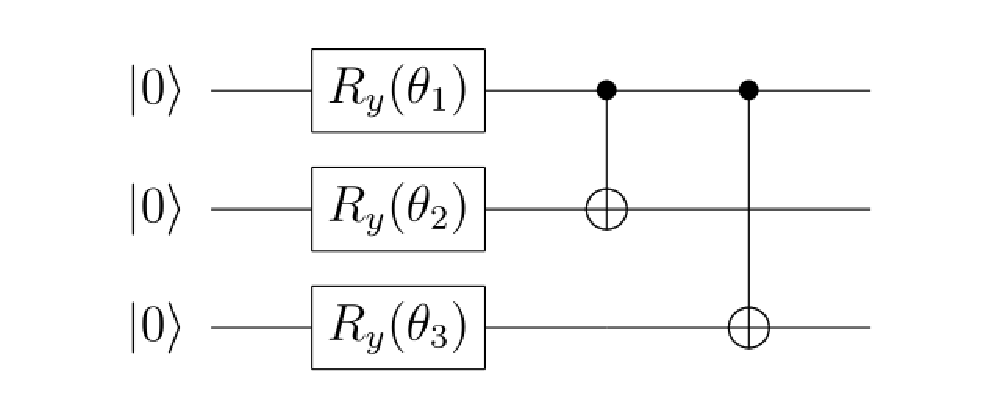}
      \caption{The quantum circuit diagram of the preparation of the initial quantum states.}
  \end{figure}

Without loss of generality, we take $\theta_{1}={\pi}/{4}$, $\theta_{2}={\pi}/{6}$, $\theta_{3}={\pi}/{8}$.
Then, we perform quantum simulation on the {\it Wuyuan 2} device to test our constraint relation.
We repeat the experiment 5 times and each circuit is implemented in 10000 shots.
The set of 4 tomographic measurement operators and the corresponding counts for the qubit \textsl{A}, \textsl{B} or \textsl{C} are respectively shown in Table 1.
\begin{table}[tbhp]
      \begin{tablenotes}
        \centering
        \footnotesize
        \item[]Table 1. Tomographic measurement operators and the corresponding counts for each qubit.
      \end{tablenotes}
\centering
\begin{tabular}{c c c c r}
\hline
\hline
$\mu_{i}$ & $n_{i}$ & qubit \textsl{A} & qubit \textsl{B} & qubit \textsl{C}\\
\hline
$\mu_{0}=|0\rangle\langle0|+|1\rangle\langle1|$ & $n_{0}$ & 5000 & 5000 & 5000 \\
\hline
$\mu_{1}=|0\rangle\langle0|$ & $n_{1}$ & 8252 & 7743 & 7740 \\
\hline
$\mu_{2}=|-\rangle\langle-|$ & $n_{2}$ & 4411 & 2729 & 3196 \\
\hline
$\mu_{3}=|R\rangle\langle R|$ & $n_{3}$ & 4989 & 4951 & 4941 \\
\hline
\hline
\end{tabular}
\end{table}

To proceed, we calculate the Stokes parameters.
Substituting these results into the formula (\ref{single_rho}) yields

\begin{eqnarray}\label{}
  \rho_{A}=
          \left(
               \begin{array}{cc}
                     0.8252& ~0.0589-\textrm{i}0.0011\\
                     0.0589+\textrm{i}0.0011& ~0.1748\\
               \end{array}
          \right),
\end{eqnarray}

\begin{eqnarray}\label{}
  \rho_{B}=
          \left(
               \begin{array}{cc}
                     0.7743& ~0.2271-\textrm{i}0.0049\\
                     0.2271+\textrm{i}0.0049& ~0.2257\\
               \end{array}
          \right)
\end{eqnarray}
and
\begin{eqnarray}\label{}
  \rho_{C}=
          \left(
               \begin{array}{cc}
                     0.7740& ~0.1804-\textrm{i}0.0059\\
                     0.1804+\textrm{i}0.0059& ~0.2260\\
               \end{array}
          \right).
\end{eqnarray}

A straightforward calculation shows that these density matrices satisfy the following properties: (\textrm{i}) the normalization, $\textrm{Tr}(\rho_k)=1$; (\textrm{ii}) the Hermiticity, $\rho^{\dag}_k=\rho_k$; and (\textrm{iii}) the positive semidefiniteness, the eigenvalues lie in the interval $[0,1]$ and $\textrm{Tr}(\rho^{2}_k)<1$, where $k=A,B,C$.

Then we calculate the quantities $V_{k}$, $P_{k}$ and $Q$, as shown in Table 2.
\begin{table}[tbhp]
\centering
  \begin{tablenotes}
    \centering
    \footnotesize
    \item[]Table 2. Experimental results of $V_{k}$, $P_{k}$ and $Q$.
  \end{tablenotes}
\begin{tabular}{c c c c c c r}
\hline
\hline
\ $V_{A}$ & $V_{B}$ & $V_{C}$ & $P_{A}$ & $P_{B}$ & $P_{C}$ & $Q$ \\
\hline
0.1178 & 0.4543 & 0.3610 & 0.6504 & 0.5486 & 0.5480 & 0.5420  \\
\hline
\hline
\end{tabular}
\end{table}

By this, one can obtain
\begin{eqnarray}\label{}
Q+\frac{1}{3}\sum_{k=A,B,C}V_{k}^{2}+\frac{1}{3}\sum_{k=A,B,C}P_{k}^{2}=1.0003.
\end{eqnarray}
The result is very close to the theoretical value 1, given by Eq. (\ref{triality}). It means that, the constraint relation of the entanglement measure, quantitative waviness and particleness can be tested by using the quantum cloud platform.

In order to provide some perspective on how the tripartite entanglement constraint on wave-particle duality, we now consider
two cases with a random variable $\theta$, e.g., $\theta_{1}=\theta_{2}=\theta_{3}=\theta$ and $\theta_{1}=\theta$, $\theta_{2}=\theta_{3}=0$ in the state (\ref{3-thete-state}).

Theoretically, for $\theta_{1}=\theta_{2}=\theta_{3}=\theta$ one gets
\begin{eqnarray}\label{}
\frac{1}{3}\sum_{k=A,B,C}V_{k}^{2}=\frac{1}{3}(\sin^{6}\theta+2\sin^{2}\theta),
\end{eqnarray}
\begin{eqnarray}\label{}
\frac{1}{3}\sum_{k=A,B,C}P_{k}^{2}=\frac{1}{3}(2\cos^{4}\theta+\cos^{2}\theta),
\end{eqnarray}
\begin{eqnarray}\label{}
Q=-\frac{1}{3}(\sin^{6}\theta+2\sin^{4}\theta-3\sin^{2}\theta);
\end{eqnarray}
and for $\theta_{2}=\theta_{3}=0$, $\theta_{1}=\theta$ we have
\begin{eqnarray}\label{}
\frac{1}{3}\sum_{k=A,B,C}V_{k}^{2}=0,
\end{eqnarray}
\begin{eqnarray}\label{}
\frac{1}{3}\sum_{k=A,B,C}P_{k}^{2}=\cos^{2}\theta,
\end{eqnarray}
\begin{eqnarray}\label{}
Q=\sin^{2}\theta.
\end{eqnarray}
Taking $\theta=0, \pi/4, \pi/2, 3\pi/4, \pi$, respectively, we perform quantum simulation on the {\it Wuyuan 2} device.
Finally, we calculate each quantity on the left hand side of the Eq. (\ref{triality}), as shown in Table 3 for $\theta_{1}=\theta_{2}=\theta_{3}=\theta$ and Table 4 for $\theta_{1}=\theta$, $\theta_{2}=\theta_{3}=0$.
\begin{table}[h]
\centering
  \begin{tablenotes}
    \centering
    \footnotesize
    \item[]Table 3. Experimental results for $\theta_{1}=\theta_{2}=\theta_{3}=\theta$.
  \end{tablenotes}
\begin{tabular}{c c c c r}
\hline
\hline
\ $\theta$ & $\frac{1}{3}\Sigma_{k=A,B,C}V_{k}^{2}$ & $\frac{1}{3}\Sigma_{k=A,B,C}P_{k}^{2}$ & $Q$ & $\textrm{sum}$  \\
\hline
0 & 0.0001 & 0.8960 & 0.1030 & 0.9991 \\
\hline
$\frac{\pi}{4}$ & 0.3322 & 0.2910 & 0.3770 & 1.0002 \\
\hline
$\frac{\pi}{2}$ & 0.8670 & 0.0001 & 0.1330 & 1.0001 \\
\hline
$\frac{3\pi}{4}$ & 0.3320 & 0.2930 & 0.3750 & 1.0000 \\
\hline
$\pi$ & 0.0001 & 0.8680 & 0.1370 & 1.0051 \\
\hline
\hline
\end{tabular}
\end{table}

\begin{table}[h]
\centering
  \begin{tablenotes}
    \centering
    \footnotesize
    \item[]Table 4. Experimental results for $\theta_{1}=\theta$, $\theta_{2}=\theta_{3}=0$.
  \end{tablenotes}
\begin{tabular}{c c c c r}
\hline
\hline
\ $\theta$ & $\frac{1}{3}\Sigma_{k=A,B,C}V_{k}^{2}$ & $\frac{1}{3}\Sigma_{k=A,B,C}P_{k}^{2}$ & $Q$ & $\textrm{sum}$  \\
\hline
0 & 0.0001 & 0.9000 & 0.0990 & 0.9991 \\
\hline
$\frac{\pi}{4}$ & 0.0002 & 0.4430 & 0.5570 & 1.0002 \\
\hline
$\frac{\pi}{2}$ & 0.0002 & 0.0001 & 0.9997 & 1.0000 \\
\hline
$\frac{3\pi}{4}$ & 0.0002 & 0.4430 & 0.5570 & 1.0002 \\
\hline
$\pi$ & 0.0002 & 0.8810 & 0.1190 & 1.0002 \\
\hline
\hline
\end{tabular}
\end{table}

As shown in Figs. 2 and 3, furthermore, we show the experimental test as well as the theoretical results, for these two cases, respectively. The solid lines are the theoretical results to fit the experimentally achieved visibility and the experimental data are respectively taken to $\theta = 0, \pi/4, \pi/2, 3\pi/4, \pi$.
As a result, it can be found that these experimental outcomes are almost identical with the theoretical values.
The difference between the experimental and theoretical results are mainly caused by the qubit crosstalk, CNOT gates in quantum circuit and readout errors, etc.
It is worth mentioning that the constraint of the Eq. (\ref{triality}) on their variation is very strong, since the sums of these three quantities (quantitative waviness, particleness and the global entanglement) are always taken to 1.

\begin{figure}[h]
      \centering
      \includegraphics[width=5in]{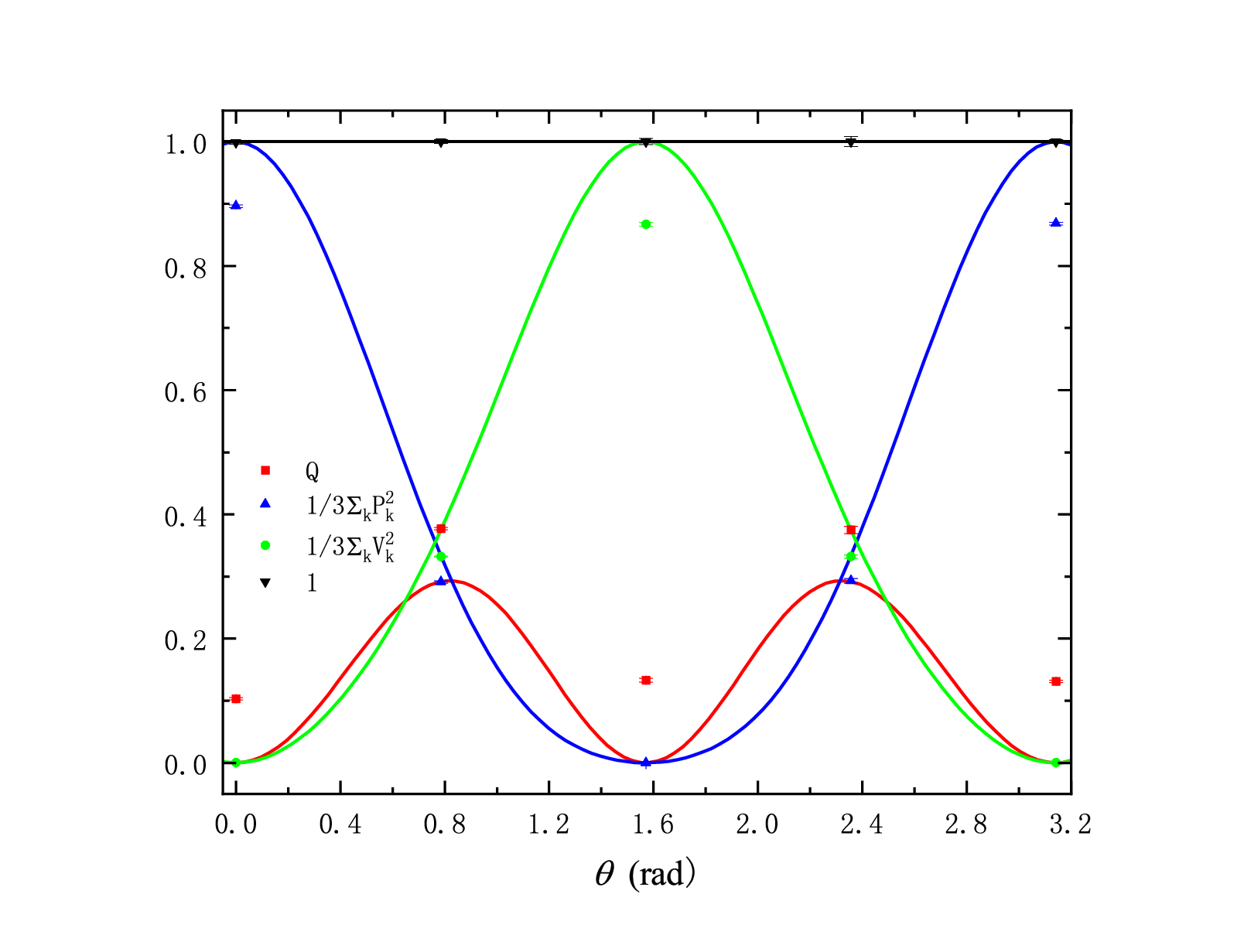}
      \caption{Theoretical results versus experimental tests for cases $\theta_{1}=\theta_{2}=\theta_{3}=\theta$.}
  \end{figure}

\begin{figure}[h]
      \centering
      \includegraphics[width=5in]{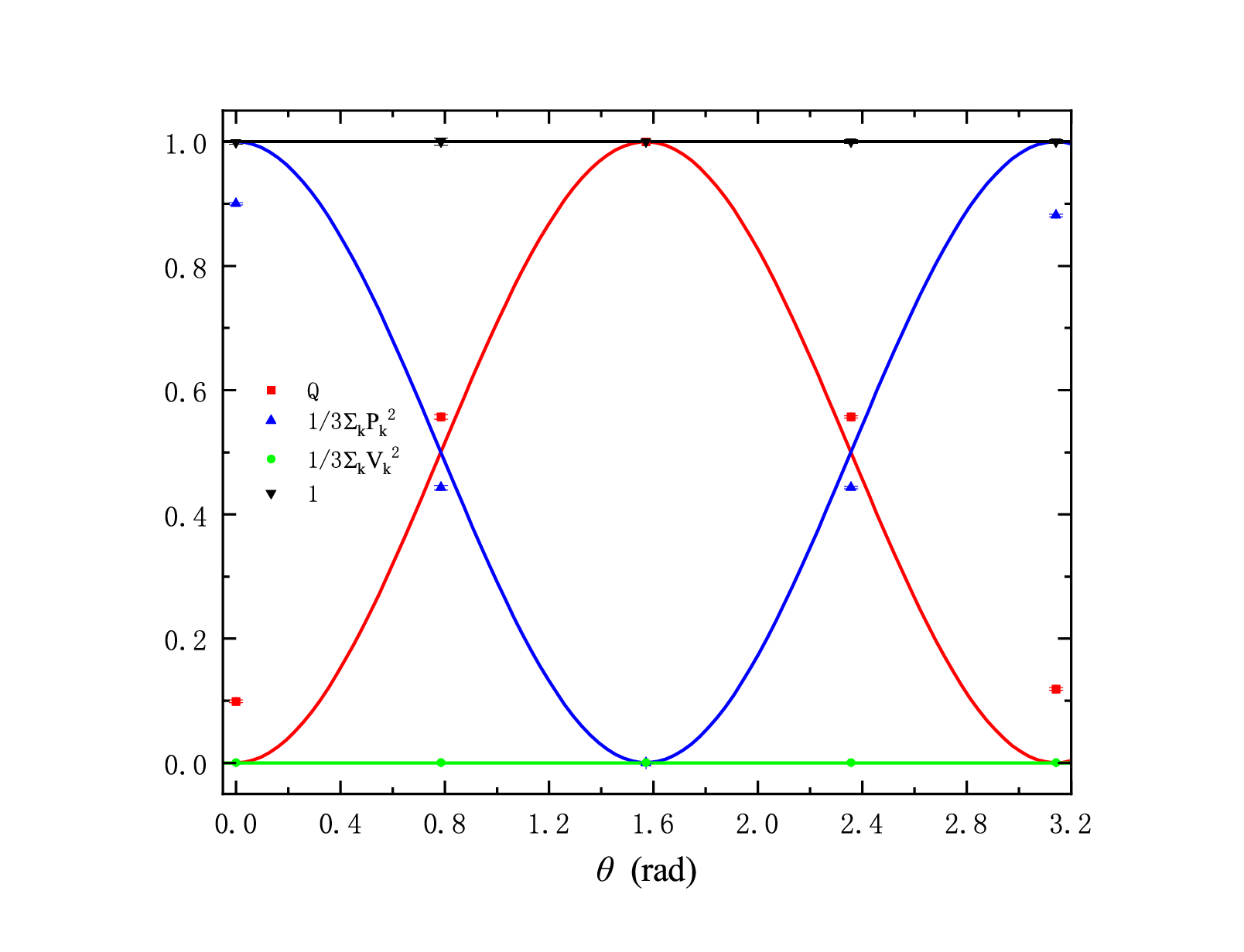}
      \caption{Theoretical results versus experimental tests for cases $\theta_{1}=\theta$, $\theta_{2}=\theta_{3}=0$.}
  \end{figure}

\section{Discussion and summary}

In summary, we have proposed a constraint relation involving the global tripartite entanglement measure and quantitative wave-particle duality for the tripartite systems. In terms of quantum state tomography, we have performed quantum simulation to test the constraint relation on the {\it Wuyuan 2} superconducting quantum device.
It is found that the results of experimental test are almost identical with the theoretical values.
This means that the quantitative waviness and particleness are closely related to the global entanglement in quantum system.

\begin{acknowledgements}

We acknowledge the use of the OriginQ Quantum Experience for this work. The views expressed are those of the authors and do not reflect the official policy or position of OriginQ or the OriginQ Quantum Experience team.
This work was supported by
the National Natural Science Foundation of China under Grant Nos: 62271189, 12071110,
the Hebei Central Guidance on Local Science and Technology Development Foundation of China under Grant Nos: 226Z0901G, 236Z7604G,
the Hebei 3-3-3 Fostering Talents Foundation of China under Grant No: A202101002,
the National Social Science Fund of China under Grant No: 23BZX103,
the Education Department of Hebei Province Natural Science Foundation of China under Grant Nos: ZD2021407, ZD2021066,
the Education Department of Hebei Province Teaching Research Foundation of China under Grant No: 2021GJJG482.

\end{acknowledgements}
\noindent{\large \bf Declarations}

{ \bf Data availability statement}

The authors confirm that the data supporting the findings of this study are available within the article.

{ \bf Funding and/or Conflicts of interests/Competing interests}

The authors declare that they have no conflicts of and competing interests.



\end{document}